\documentstyle[12pt]{article}
\def\lapproxeq{\lower .7ex\hbox{$\;\stackrel{\textstyle <}{\sim}\;$}}
\def\gapproxeq{\lower .7ex\hbox{$\;\stackrel{\textstyle >}{\sim}\;$}}

\begin{document}

\titlepage

\begin{flushright} RAL-93-071 \\ October 1993
\end{flushright}

\begin{center}
\vspace*{2cm}
{\large{\bf Determining the gluon distributions in the proton and photon  from
two-jet production at HERA}}
\end{center}

\vspace*{.75cm}
\begin{center}
J.R.\ Forshaw and R.G.\ Roberts \\
Rutherford Appleton Laboratory, \\ Chilton, Didcot OX11 0QX, England. \\
\end{center}

\vspace*{1.5cm}

\begin{abstract}
Two-jet production from the direct photon contribution at HERA is a
sensitive measure of the small-$x$ gluon in the proton. We propose
measurements of ratios of the jet cross-sections which will clearly
distinguish between gluons with or without singular behaviour at small $x$.
Furthermore, we show that analogous ratio measurements for the resolved
photon contribution provide a sensitive way of determining the gluon
distribution in the photon.
\end{abstract}

\newpage

{\large {\bf Introduction}}
\vspace*{0.5cm}

One of the aims of HERA is to study the structure of the proton and in
particular to learn about the distribution of gluons at small $x$ (i.e. $x$
\lapproxeq 10$^{-2}$). Here the gluon is expected to be dominant and we are in
an unexplored region of lepton-hadron scattering. The latest measurements of
the proton structure function $F_2^p(x,Q^2)$ from H1 \cite{h1f2} and ZEUS
\cite{zeusf2} show the distinct rise for $x \lapproxeq 10^{-2}$ which has
encouraged speculation that this may be a signal of so-called `Lipatov'
behaviour \cite{bfkl}. On the other hand it is possible to generate such a
rise using conventional leading log $Q^2$ evolution from a `valencelike' gluon
distribution at $Q_0^2=0.3$ GeV$^2$ \cite{grvprot}. Measuring $F_2^p$ however
is only an {\it indirect} probe of the gluon and this is the main reason why
the gluon is the least constrained of all the parton distributions to date.
There have been several methods proposed which aim to extract the small-$x$
gluon content of the proton at HERA rather more directly \cite{amcs,mns,bk}.

Here, we present an analysis of two-jet photoproduction which utilises the
separation of events into so-called `direct' and `resolved' components. Through
the construction of appropriate cross section ratios we show that HERA
should provide a sensitive and direct measurement of the small-$x$ gluon in
the proton in addition to revealing important information regarding the gluon
in the photon.

The first step we take is analogous to the procedure previously considered
for hadronic two-jet production \cite{mrs2j,cdf2j} where the configuration of
`same-side' jets allows very small $x$ values of the initial partons to be
examined. At HERA we propose isolating the contribution to jet photoproduction
from `direct' photon interactions (see fig.1(a)). We focus on the region where
both jets have equal rapidities and are travelling in the electron direction
(which we define to be that of positive rapidity). Such large and positive
rapidities mean that we are sensitive to the small-$x$ gluon within the proton.

The second step is to consider contributions like those of fig.1(b), where the
photon is resolved into its partonic components. We now study the region of
negative jet rapidities where, conversely to the first step, the cross
section is sensitive to the parton distributions in the photon for values of
$x_{\gamma}$ around $0.2 - 0.5$. With the proton gluon distribution
already constrained from step one, the resolved photon contribution to
two-jet production provides a sensitive measure of the parton distributions in
the photon. The possibility of using the two-jet data from HERA to provide
information on the parton content of the photon, especially on the gluon
distribution has been emphasised by several groups \cite{gs,dg,boo}. We wish
to emphasise that measurement of {\it ratios} of the two-jet cross sections is
a particularly clean and sensitive way of discriminating between different
parametrisations of the gluon distribution which tends to minimise the
experimental and theoretical uncertainties involved.

\vspace*{1.0cm}
{\large {\bf Step 1 $-$ the gluon in the proton}}
\vspace*{0.5cm}

To begin with, consider the direct photon contribution to two-jet production,
e.g. see fig.1(a). The two-jet cross section may be written to leading order
(LO) as

\begin{equation}
\frac{d^3\sigma^{dir}}{dy_1dy_2dp^2_T} \; = \;
zf_{\gamma /e}(z)\; xf_{j/p}(x,\mu)\; \frac{d \hat \sigma(\gamma j \rightarrow
k \l)} {d \hat t}  \label{d2sigdir}
\end{equation}

\noindent
where $f_{\gamma /e}(z)$ is the probability that the incoming electron will
emit an effectively real photon with a fraction $z$ of its momentum and
$f_{j/p}(x,\mu)$ is the momentum distribution of parton $j$ inside the proton
evaluated at scale $\mu$. The laboratory rapidities ($y_1$ and $y_2$) of the
final state partons (emitted with transverse momentum $p_T$) can be expressed
in terms of the momentum fractions $z$ and $x$:
\begin{eqnarray}
z = \frac{p_T}{\sqrt s}\; \sqrt {\frac{E_p}{E_e}}\; (e^{y_1} + e^{y_2})
\nonumber \\
x = \frac{p_T}{\sqrt s}\; \sqrt {\frac{E_e}{E_p}}\; (e^{-y_1} + e^{-y_2}).
\label{xzdefn}
\end{eqnarray}
The energies of the incoming electron and proton are $E_e$ and $E_p$
respectively and the centre-of-mass energy is $\sqrt s = 2\sqrt{E_e E_p}$.
In this analysis we shall study only the configuration where the two jets have
equal rapidities, i.e. $y_1 = y_2 = y$. At HERA energies, choosing $y \sim 1$
and $p_T = 5$ GeV, we find $x \simeq .003$.

There are two subprocesses in the direct photon contribution: (i) $\gamma \;
g \rightarrow q \; \bar q$ and (ii) $\gamma \; q (\bar q) \rightarrow
g \;q (\bar q)$ and the LO cross section of eqn.(\ref{d2sigdir}) takes the
simple form
\begin{equation}
\left . \frac{d^3\sigma^{dir}}{dy_1dy_2dp^2_T} \right |_{y_1=y_2=y} \;=\;
\;\frac{\pi \alpha \alpha_s}{4p_T^4}\;zf_{\gamma/e}(z)\;
\left [(\sum_{i} e_i^2)xg(x,\mu)\;+\;(10/3)F_2^p(x,\mu)\right ].
\label{xsecn}
\end{equation}
Since $xg(x,\mu)$ is much larger than $x\bar q(x,\mu)$ at small $x$ then (i)
dominates the cross section for $x \lapproxeq 10^{-2}$. Thus evaluating ratios
of the above cross section at different values of $y$ effectively measures the
ratio of the gluon distribution at different values of $x$. We suggest
measuring the ratio $\sigma_{dir}(y)/\sigma_{dir}(y=-0.5)$ (where we have
integrated over all $p_T \ge 5$ GeV in eqn.(\ref{xsecn}) to construct
$\sigma_{dir}(y)$). Fig.2 shows the expectations for this ratio from various
proposed gluons: $MRS D0'$, $MRS D-'$\cite{mrs} and $GRV$\cite{grvprot}. The
$D-'$ curve is generated from a singular gluon which, at $\mu ^2 = 4$ GeV$^2$
and small $x$, behaves as $ xg(x,\mu) \sim x^{-1/2}$ which is in contrast to
the $D0'$ curve which corresponds to $xg(x,\mu)\rightarrow$ constant as $x
\rightarrow 0$ at $\mu ^2 = 4$ GeV$^2$. The $GRV$ prediction leads to a
singular gluon distribution for $\mu^2 \gapproxeq 1$ GeV$^2$ due to the long
evolution time which results from starting evolution at low scales.
The curves of fig.2 were computed with $z$ in the range $0.2 \le z \le 0.7$.
These are typical HERA cuts and are intended to reduce background uncertainties
due to beam gas interactions ($z > 0.2$) and DIS interactions ($z < 0.7$), they
are also responsible for the cusps around $y = 0.1$.

The scale $\mu$ was taken to be $p_T/2$, as specified
by the prescription of Ellis, Kunszt and Soper \cite{eks} who have shown that
the scale $\mu$ for which the leading order calculation reproduces the less
scale dependent ${\cal O}(\alpha_s^3)$ result is given by:
\begin{equation}
\mu \approx \frac{{\rm cosh}(y_1-y_2)}{{\rm cosh}(0.7(y_1-y_2))} \frac{p_T}{2}
\quad \stackrel{y_1=y_2}{\longrightarrow} \quad \frac{p_T}{2}.
\label{mu}
\end{equation}
This (approximate) equality of the LO and NLO cross sections for the choice of
a common scale $\mu = {1\over 2}p_T$ when $y_1=y_2$ is also consistent with
the results of B\"odecker for two-jet direct photoproduction \cite{bod}.
Varying the value of $\mu$, e.g. to $p_T$, does not seriously alter the curves
in fig.2.

As the figure demonstrates, this is a sensitive measure since it effectively
measures the ratio of the magnitude of the gluon at small $x$ to that
around $x \sim 0.1$. The ratio differs by as much as 40\% between the
$MRS D0'$ and $MRS D-'$ predictions and experiment should easily be able to
distinguish between them. Note that we choose to evaluate the cross sections at
$y_1 = y_2$ since this maximises the sensitivity to the particular choice of
gluon distribution, otherwise contributions from larger $x$ values
tend to smear out the small-$x$ contribution.

Of course the predictions we have for the ratio in fig.2 does assume that
one can indeed separate the direct and resolved components of the photon on an
event by event basis. Recent preliminary results from HERA demonstrate that
making a suitable cut on $x_\gamma$ (see fig.1(b)), e.g $x_\gamma \ge 0.75$,
can provide the required clean separation \cite{marseille}. Furthermore, we
have checked, in LO, that the direct cross section is almost unaltered after
including the resolved component with $x_{\gamma} \ge 0.75$.

\vspace*{1.0cm}
{\large {\bf Step 2 $-$ the gluon in the photon}}
\vspace*{0.5cm}

We next turn to the resolved contribution, e.g. fig.1(b), where the two-jet
cross section, to leading order, is given by

\begin{equation}
\frac{d^3\sigma^{res}}{dy_1dy_2dp^2_T} \; = \;\int dz \, f_{\gamma /e}(z)\;
x_\gamma f_{i/\gamma}(x_\gamma,\mu) \; xf_{j/p}(x,\mu)\;
\frac{d \hat \sigma(i j  \rightarrow k \l)} {d \hat t}
\label{d2sigres}
\end{equation}
and the expression for $z$ in eqn.(\ref{xzdefn}) is now the expression for
the product $zx_\gamma$.

In the previous section, we saw that choosing $y_1$, $y_2$ large and positive
led to small $x$ for the proton, now we see that choosing $y_1$,
$y_2$ large and negative leads to small $zx_\gamma$. If we compute the
average  values of $x_\gamma$ we find that typically we are exploring the
range $0.18 < x_\gamma < 0.5$ for $-2 < y_1=y_2 < 0$.

Instead of having only two subprocesses, as in step one, we now have the
full set of leading order graphs which contribute to jet cross sections
in hadron-hadron scattering. We will assume that, from the analysis in step
one, the small-$x$ behaviour of the partons in the proton is now sufficiently
pinned down so that an analysis of the two-jet cross section can yield
direct information on the partons in the photon. Again we take the scale
$\mu={1\over 2}p_T$ in the LO expressions for hard cross sections.
To a large extent, the quark distributions in the photon are
already constrained from fits to $F_2^\gamma$ and this is reflected in various
popular parametrisations \cite{gs,dg,grv,lac} where the major differences
all reside in the assumed behaviour of the gluon.

To emphasise the role of the resolved contribution we therefore shift towards
more negative values of the rapidities $y_1$ and $y_2$. In fig.3 we show
predictions for the ratio of the resolved photon cross section (defined
as the component with $x_\gamma < 0.75$) for $-2 \le y \le 0$ to the direct
photon two-jet cross section at $y=0$. We choose to normalise to the direct
cross section at $y=0$ since this choice gives good sensitivity to the
normalisation (as well as the shape) of the parton distributions in the photon,
as well as ensuring that good statistics should be possible ($y=0$ is near the
peak of the direct cross section rapidity distribution). Again we integrate
over $p_T \ge 5$ GeV. It is clear from the figure that the large variations
imply that the ratio is indeed a sensitive measure of the gluon density in the
photon.

\vspace*{1.0cm}
{\large {\bf Conclusions}}
\vspace{0.5cm}

We have proposed that a measurement of the two-jet cross section ratio,
$\sigma_{dir}(y) /\sigma_{dir}(-1)$, for $y_1=y_2=y$ from
photoproduction at HERA will provide a sensitive test of the gluon
density in the proton. The relatively large values of $y$ which are accessible
enable the gluon content of the proton to be probed in the interesting
small-$x$ region. Since we expect many of the experimental uncertainties
involved in computing the ratio to cancel (and that any rapidity dependence of
the detector acceptance can be corrected for) we expect that the construction
of such a cross section ratio will be experimentally straightforward.

Already, events due to hard two-jet production at HERA have been seen clearly
for $p_T$ up to 20 GeV \cite{marseille} $-$ even a separation of these events
into `enriched direct $\gamma$' and `enriched resolved $\gamma$' samples
has been performed. These observations are extremely encouraging for the
method we suggest since, in step one, the small-$x$ gluon in the proton is
determined from the `clean' direct photon component. Having pinned down the
gluon in the proton one can then proceed to extract information on the gluon
density in the photon. This is step two of our recipe in which only the
resolved component of the photon contributes and where a measurement of the
ratio, $\sigma_{res}(y)/\sigma_{dir}(y=0)$, can go a long way in discriminating
between the various proposed gluon distributions of the photon.

We believe that using {\it ratios} of the cross sections in this way reduces
both experimental and theoretical uncertainties and allows a relatively
simple and logical procedure to provide a sensitive determination of the gluon
densities in both the proton and photon.

\vspace*{1cm}
\noindent {\large {\bf Acknowledgements}}
\vspace*{0.5cm}

It is a pleasure to thank Jonathan Butterworth for many helpful discussions
and suggestions.

\newpage

\vskip 1.0cm
\noindent{\large\bf Figure Captions}

\begin{itemize}
\item[Fig.\ 1] (a) A direct photon contribution to the hard two-jet
cross section. \newline \indent (b) A resolved photon contribution to the hard
two-jet cross section.

\item[Fig.\ 2] The ratio
$\left . \frac{d^2\sigma^{dir}}{dy_1dy_2} \right |_{y_1=y_2=y}
/\left . \frac{d^2\sigma^{dir}}{dy_1dy_2} \right |_{y_1=y_2=-0.5}$
for three choices of proton

\vspace*{0.2cm}
gluon density, i.e. $MRS D0'$, $MRS D-'$
\cite{mrs} and $GRV$\cite{grvprot}.

\item[Fig.\ 3] The ratio
$\left . \frac{d^2\sigma^{res}}{dy_1dy_2} \right |_{y_1=y_2=y}
/\left . \frac{d^2\sigma^{dir}}{dy_1dy_2} \right |_{y_1=y_2=0}$
  using (a) $MRS D-'$ and (b) $MRS D0'$ for the proton parton densities.
The labelled curves correspond to various choices of the photon parton
densities, LAC1, LAC3 \cite{lac}, GS2 \cite{gs}, GRV \cite{grv} and
DG \cite{dg}. Also shown are the corresponding predictions when the
gluon density in the photon is switched off.

\end{itemize}


\newpage

\begin{thebibliography}{999}
\bibitem{h1f2} H1 Collaboration, I.\ Abt {\it et al}., DESY report
DESY 93-117, Aug. 1993.
\bibitem{zeusf2} ZEUS Collaboration, M.\ Klein {\it et al}., DESY report
DESY 93-110, Aug. 1993.
\bibitem{bfkl} E.A.\ Kuraev, L.N.\ Lipatov and V.S.\ Fadin, Sov. Phys.
JETP {\bf 45} (1977) 199; \newline
Ya.\ Ya.\ Balitsky and L.N.\ Lipatov, Sov. J. Nucl. Phys. {\bf 28} (1978)
822.
\bibitem{grvprot} M.\ Gl\"uck, E.\ Reya and A. Vogt, Phys. Lett. {\bf B306}
(1993) 391.
\bibitem{amcs} A.M.\ Cooper-Sarkar {\it et al}., Zeit. Phys. {\bf C39}
(1988) 281.
\bibitem{mns} A.D.\ Martin, C.K.\ Ng and W.J.\ Stirling, Phys.
Lett. {\bf B191} (1987) 200.
\bibitem{bk} J.\ Bl\"umlein and M.\ Klein, DESY report DESY 92-038 (1992);
\newline K.\ Prytz, Phys. Lett. {\bf B311} (1993) 286.
\bibitem{mrs2j}A.D.\ Martin, W.J.\ Stirling and R.G.\ Roberts, RAL
preprint-93-047 (1993).
\bibitem{cdf2j}CDF collaboration: contribution to the EPS Conference on High
Energy Physics, Marseilles, July 1993, preprint FERMILAB-Conf-93/203-E.
\bibitem{gs} L.E.\ Gordon and J.K.\ Storrow, Manchester University preprint
M/C.TH.92/06 (1992).
\bibitem{dg} M.\ Drees and R.\ Godbole, Phys. Rev. Lett. {\bf 61} (1988)
682; Phys. Rev. {\bf D39} (1989) 169.
\bibitem{boo} H.\ Baer, J.\ Ohnemus and J.F.\ Owens, Zeit. Phys. {\bf C 42}
(1988) 657.
\bibitem{mrs}A.D.\ Martin, W.J.\ Stirling and R.G.\ Roberts, Phys.\ Lett.\
{\bf 306B} (1993) 145.
\bibitem{eks} S.D\ Ellis, Z.\ Kunszt and D.E.\ Soper, Phys. Rev. Lett.
{\bf 69} (1992) 1496.
\bibitem{bod} D. B\"odecker, Zeit. Phys. {\bf C59} (1993) 501.
\bibitem{marseille} R.\ Klanner and A.\ de Roeck, HERA results presented
at the EPS conference on high energy physics, Marseille 1993.
\bibitem{grv} M.\ Gl\"uck, E.\ Reya and A. Vogt, Zeit. Phys. {\bf C53}
(1992) 127.
\bibitem{lac} H.\ Abramowicz, K.\ Charchula and A.\ Levy, Phys. Lett.
{\bf B269} (1991) 458.
\end{thebibliography}
\end{document}